\documentclass[usenatbib]{mn2e}
\usepackage{graphicx}
\usepackage{amsmath}

\def\apj{\rm ApJ}
\def\apjl{\rm ApJL}

\def\mnras{\rm MNRAS}

\def\aap{\rm AAP}

\def\gax{\mathrel{\raise.3ex\hbox{$>$}\mkern-14mu\lower0.6ex\hbox{$\sim$}}}
\def\lax{\mathrel{\raise.3ex\hbox{$<$}\mkern-14mu\lower0.6ex\hbox{$\sim$}}}
\def\gtorder{\mathrel{\raise.3ex\hbox{$>$}\mkern-14mu
             \lower0.6ex\hbox{$\sim$}}}
\def\ltorder{\mathrel{\raise.3ex\hbox{$<$}\mkern-14mu
             \lower0.6ex\hbox{$\sim$}}}

\begin{document}

\title [How the Angular Tail Wags the Radial Dog]
   {Over-constrained Models of Time Delay Lenses Redux: How the Angular Tail Wags the Radial Dog}

\author[C.~S. Kochanek]{ 
    C.~S. Kochanek$^{1,2}$ 
    \\
  $^{1}$ Department of Astronomy, The Ohio State University, 140 West 18th Avenue, Columbus OH 43210 \\
  $^{2}$ Center for Cosmology and AstroParticle Physics, The Ohio State University,
    191 W. Woodruff Avenue, Columbus OH 43210 \\
   }

\maketitle

\begin{abstract}
The two properties of the radial mass distribution of a gravitational lens that are well-constrained by
Einstein rings are the Einstein radius
$R_E$ and $\xi_2 = R_E \alpha''(R_E)/(1-\kappa_E)$, where $\alpha''(R_E)$ and
$\kappa_E$ are the second derivative of the deflection profile and the convergence
at $R_E$.  However, if there is a tight mathematical relationship between the 
radial mass profile and the angular structure, as is true of ellipsoids, an Einstein 
ring can appear to strongly distinguish radial mass distributions with the same 
$\xi_2$. This problem is beautifully illustrated by the ellipsoidal models 
in \cite{Millon2019}.  When using Einstein rings to constrain the radial mass
distribution, the angular structure of the models must contain all the degrees
of freedom expected in nature (e.g., external shear, different ellipticities for the stars and the
dark matter, modest deviations from elliptical structure, modest twists of the
axes, modest ellipticity gradients, etc.) that work to decouple the radial and
angular structure of the gravity.  Models of Einstein rings with too few
angular degrees of freedom will lead to strongly biased likelihood
distinctions between radial mass distributions and very precise but inaccurate 
estimates of $H_0$ based on gravitational lens time delays.  
\end{abstract}


\section{Introduction}
\label{sec:introduction}

Measurements of $H_0$ from time delays scale suffer from the degeneracy that
$H_0 \propto 1 - \kappa_E$ (\citealt{Kochanek2002}, \citealt{Kochanek2006})
where a fundamental mathematical degeneracy means
that no differential lens data (positions, fluxes, etc.) other than
time delays can determine the
convergence $\kappa_E$ at the Einstein radius (see, e.g., \citealt{Gorenstein1988},
\citealt{Kochanek2002}, \citealt{Kochanek2006}, \citealt{Schneider2013},
\citealt{Wertz2018}, \citealt{Sonnenfeld2018}, \citealt{Kochanek2019}).  The properties of the radial mass distribution
that are determined by such data are the Einstein radius $R_E$ and the
dimensionless quantity $\xi_2 = R_E \alpha''(R_E)/(1-\kappa_E)$ where
$\alpha''(R_E)$ is the second derivative of the deflection profile at
$R_E$ (\citealt{Kochanek2019}).  The mathematical structure of the mass
model then determines $\kappa_E$ given the available constraints on
$R_E$ and $\xi_2$ and the amount of freedom in the mass model.

The two parameter, or effectively two parameter, mass models that are in
common use lead to a unique value for $\kappa_E$ given $R_E$ and $\xi_2$.
For example, the power law model with $\alpha(r) = b^{n-1} r^{2-n}$
has $R_E=b$, $\xi_2=2(n-2)$ and $\kappa_E = (3-n)/2 = (2-\xi_2)/4$.
While it is frequently said that lenses prefer density distributions
similar to the singular isothermal sphere with $n \simeq 2$
(e.g., \citealt{Rusin2005}, \citealt{Gavazzi2007}, \citealt{Koopmans2009},
\citealt{Auger2010}, \citealt{Bolton2012}), the real constraint is
that $\xi_2 \simeq 0$, which the power law models produce
for $n \equiv 2$.  This property makes it very dangerous to use lensing
data that strongly constrain $\xi_2$ in mass models with too few
degrees of freedom because they force the model to a particular
value of $\kappa_E$ and an estimate $H_0$ that is very precise but
potentially inaccurate.

In \cite{Kochanek2019}, we extensively demonstrate these
points and find that the accuracy of present estimates of $H_0$
from lens time delays is likely $\sim 5\%$ regardless of the
reported precision of the measurements.  The only way to avoid this
problem is to use mass models with more degrees of freedom so that
the relationship between $\xi_2$ and $\kappa_E$ is not one-to-one,
with the obvious consequence of larger uncertainties.  Since the
fundamental problem is related to systematic uncertainties in the
structure of galaxies and their dark matter halos, averaging results
from multiple lenses will not necessarily lead to any improvements
in the accuracy.

Recently, \cite{Millon2019} presented a rebuttal of \cite{Kochanek2019},
on two levels.  First, they argue that their ``black box'' lens modeling 
by its very complexity and sophistication must clearly outperform 
``toy models.''  Second, they present a series of illustrative models
that appear to strongly distinguish different radial mass distributions
through differences in the goodness of fit.  In practice, \cite{Millon2019}
actually provides a beautiful example of the consequences
of using over-constrained lens models, albeit as a case where having
too few degrees of freedom in the angular structure of the mass distribution
leads to an apparent, but illusory, ability to distinguish radial mass 
distributions at very high statistical significance.  

In this paper we use mass distributions designed to mimic those in
\cite{Millon2019} to illustrate these two points.  First, in \S2 we
discuss the problem using simple analytic models.
Then in \S3, we show
that Einstein ring data does not contain the information needed to distinguish 
radial mass distributions with the same $\xi_2$ at high statistical 
significance.  In \S4 we show that the assumption that the 
angular mass distribution is simply an ellipsoid 
distinguishes the radial mass distributions with the enormous 
statistical significances found by \cite{Millon2019}, but that 
this apparent statistical power to distinguish between radial
mass distributions vanishes as more angular degrees of freedom
are added to the model.  We summarize the results in \S5.

\section{Simple Theoretical Considerations}
\label{sec:theory}

In this section we first briefly review the discussion of 
constraints on the radial (monopole) mass distribution
of lenses, but the main focus will be on exploring the 
mathematics of the constraints associated with the quadrupole
($m=2$) of the mass distribution.  Where we need an
analytic model we will used the softened power-law mass
distribution, which we will also use in the later numerical
experiments since it is the model at the center of the
\cite{Millon2019} numerical experiments.  The model has 
convergence  
\begin{equation}
   \kappa(r) = {3-n \over 2} { A \over \left(r^2+s^2 \right)^{(n-1)/2}}
\end{equation}
and deflection profile
\begin{equation}
   \alpha(r) = { A \over r} 
   \left[ \left(r^2+s^2\right)^{(3-n)/2} -s^{3-n} \right]
\end{equation}
when circular.
In the limit of a singular model ($s\equiv0$), these become
$\kappa(r) = (3-n)A r^{1-n}/2$ and
$\alpha(r) =  A r^{2-n}$, respectively.  For these singular
cases, the normalization factor is related to the Einstein
radius by $ A = R_E^{n-1}$, to give the more familiar
forms of $\kappa(r) = (3-n) R_E^{n-1} r^{1-n}/2$ and
$\alpha = R_E^{n-1} r^{2-n}$.  We use the normalization constant
$A$ for the general case with a core radius because it is no  longer
trivially related to $R_E$.  

\begin{figure*}
\centering
\includegraphics[width=0.80\textwidth]{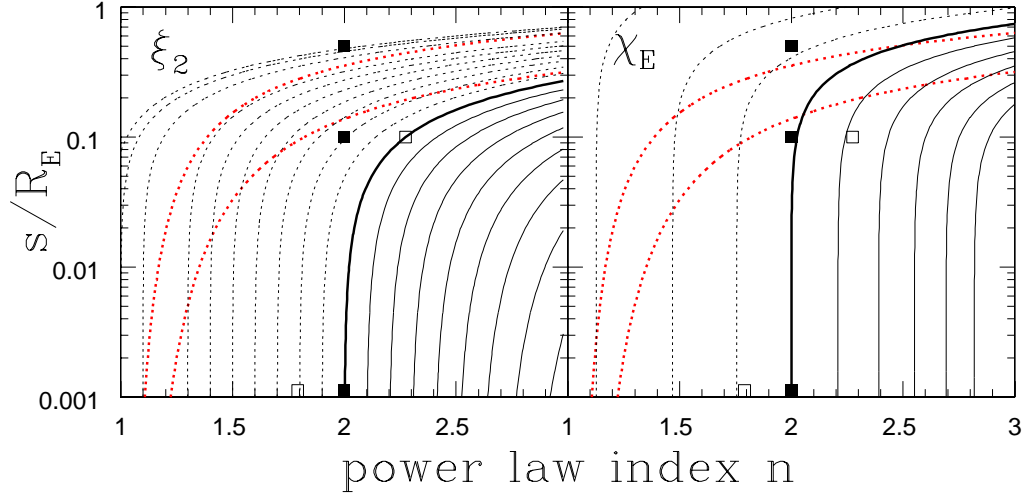}
\caption{
  The parameter space of the softened power law models. The 
  left panel shows contours of $\xi_2$ with $\xi_2=0$ for 
  the heavy solid curve and increasing (decreasing) in steps 
  of $0.2$ for the solid (dashed) contours.  
  The right panel shows contours of $\chi_E$ with $\chi_E=1/3$
  for the heavy solid curve and increasing (decreasing) in steps 
  of $0.1$ for the solid (dashed) contours.  Above
  the upper (lower) red dotted curves, the central magnification
  exceeds $1$ ($0.1$).  The solid (open) squares show some of the
  input (matched in $\xi_2$) models considered in the later
  sections. The singular models with $n=0$, $n=2$ and $n=3$
  correspond to a constant surface density, an SIS and a 
  point mass, respectively.  
  }
\label{fig:models}
\end{figure*}

Shallow power-law density profiles, power-law models with finite cores, 
\cite{Hernquist1990} and NFW (\citealt{Navarro1997}) profiles all 
produce unobserved central, ``odd'' images.  Like \cite{Millon2019}
we will simply ignore their existence in making the model
comparisons.  Fig.~\ref{fig:models} shows where the softened power
law models produce central magnifications $>0.1$ and $>1.0$ as 
a function of $n$ and $s/R_E$. The \cite{Millon2019}
model with $n=2$ and $s=0.5R_E$ is somewhat perverse because the
central image is actually magnified rather than demagnified.
We will include a model with $n=2$ and $s=0.1R_E$ that is more
reasonable (central $M<0.1$), although it would still produce a 
visible central image in many lenses.

The key point in \cite{Kochanek2019} is that the only property
of the radial mass distribution strongly constrained by lens data is
\begin{equation}
    \xi_2 = { R_E \alpha'' (R_E) \over 1 - \kappa_E }
\end{equation}
where $R_E$ is the Einstein radius, $\kappa_E$ is the convergence
and $\alpha''(R_E)$ is the second derivative of the deflection profile both
measured at the Einstein radius.  This simply comes from carrying out a Taylor
expansion of the lens equations and extracting the first term in the monopole
beyond the Einstein radius that can be constrained by lens data and expressing
it in a form which is invariant under the mass sheet degeneracy.
We have added the subscript $2$ to indicate that it is the second order term
in the expansion.  This point is independent of any angular structure
in the lens. 

It is a Taylor expansion, so data can in theory
constrain higher order terms in the structure of the monopole. The
next dimensionless, mass-sheet invariant term would be
$\xi_3 = R_E^2 \alpha'''(R_E)/(1-\kappa_e)$.  For an
Einstein radius of $R_E=1\farcs0$ and data in an annulus
$|r-R_E|/R_E = 30\%$ around $R_E$, the magnitude of the deflections 
created by $\xi_2$ are $\sim \xi_2 |r-R_E|^2/2 R_E \simeq 0\farcs045\xi_2$ 
which is relatively easy to constrain given the $0\farcs1$ resolution of
the Hubble Space Telescope (HST).  The scale of the deflections
created by $\xi_3$ are of order $\xi_3 |r-R_E|^3/6 R_E^2 \simeq 0\farcs0045\xi_3$, 
which will be difficult to constrain given both the resolution
of the data and the many systematic issues that begin to enter
on these scales (PSF models, pixelization, lens galaxy contamination,
millilensing, etc.).  
To make this a little more concrete, simply consider the power law
models where $H_0 \propto 1-\kappa_E = (2+\xi_2)/4$.  For $\kappa_E \simeq 0.5$, a 2\% 
uncertainty in $H_0$ requires a $\Delta \xi_2 \simeq 1\%$ uncertainty in $\xi_2$, 
which corresponds to deflection differences between the models
across the annulus of order $\Delta \xi_2 |r-R_E|^2/2 R_E \simeq 0\farcs00045$!   
And this is for a model which, unlike realistic models, has a one-to-one 
relation between $\kappa_E$ and $\xi_2$.

For a power law model, $\xi_2 = 2(n-2)$, so it is zero for the
$n=2$ SIS.  There is a general analytic expression for $\xi_2$
in the power law models, but it is too long to be worth
reporting.  Fig.~\ref{fig:models} shows contours of $\xi_2$ in
the $n$ and $s/R_E$ plane.  For singular models, 
$n=1$ is a constant density sheet, $n=2$ is the SIS, and $n=3$
is a point mass.  For sufficiently small cores, $\xi_2$
converges to the power law limit.  Its value decreases with 
increasing core radius at fixed exponent $n$ and increases with 
increasing $n$ at fixed core radius $s/R_E$.  Solid points mark three
of the input models we will consider ($n=2$, $s/R_E=0$, $0.1$
and $0.5$).  Open points mark the $s/R_E=0.1$ and $s/R_E=0$
models that match the values of $\xi_2$ for the first two cases.
Matching the $\xi_2$ value of the $n=0$, $s/R_E=0.5$ model requires 
power-law profile with $n<1$, which means that the surface density 
is increasing  with radius and the corresponding open point lies
off the figure to the left.  As we will see in \S3, even with 
large numbers of constraints spread across a fairly broad 
annulus, circular lens models have tremendous difficulty distinguishing
these $\xi_2$-matched models.

The simplest way to think about angular structure is in terms of 
multipoles (see \citealt{Kochanek2006}).  
For pedagogic purposes we will consider only ellipsoids
and shears (anti)aligned with the coordinate axes, although any
result can be generalized.  We consider a density distribution
$\kappa(\xi)$ with $\xi^2= x^2+y^2/q^2$ and $\epsilon=1-q$.  
For simple analytic results, we will assume $\epsilon$ is small
and expand results only to their lowest order in $\epsilon$.
If we just keep the lowest order monopole and quadrupole terms, the
monopole density is 
\begin{equation}
  \kappa_0(r) = { 1 \over 2 \pi} \int_0^{2\pi} d\theta \kappa(\xi)
    \simeq \kappa(r)
\end{equation}
and the quadrupole density is 
\begin{equation}
  \kappa_2(r) = { 1 \over \pi} \int_0^{2\pi} d\theta \cos(2\theta) \kappa(\xi)
    \simeq - \epsilon r k'(r)/2.
\end{equation}
where the limiting cases assume $\epsilon$ is small.
The combined density is 
$\kappa(r,\theta) = \kappa_0(r) + \kappa_2(r)\cos 2\theta$
which corresponds to a lensing potential of 
$\Psi(r,\theta) = \Psi_0(r) + \Psi_2(r)\cos2\theta$
where the monopole potential is
\begin{equation}
  \Psi_0 = 2 \log(r) \int_0^r du \kappa_0(u) u +
          2 \int_r^\infty du \kappa_0(u) u \ln u.
\end{equation}
We can write the quadrupole potential as
\begin{equation}
  \Psi_2 = - { 1 \over 2 } r^2 \gamma(r)
           - { 1 \over 2 } r^2 \Gamma(r)
\end{equation}
where 
\begin{equation}
   \gamma(r) = \int_r^\infty du \kappa_2(u)/u
\end{equation}
is the contribution from outside radius $r$ (i.e., like
an external shear) and
\begin{equation}
   \Gamma(r) = {1 \over r^4} \int_0^r du u^3 \kappa_2(u)
\end{equation}
is the contribution from the material inside radius $r$
(the ``internal'' shear).  Like an external shear, both 
$\gamma(r)$ and $\Gamma(r)$ are dimensionless.  The deflections 
due to the quadrupole are then
\begin{equation}
  \vec{\alpha}_2 = 
    - r \gamma(r) 
         \left[ \begin{array}{r} \cos\theta \\ -\sin\theta \end{array} \right] 
    -  r \Gamma(r) 
         \left[ \begin{array}{r} \cos3\theta \\ \sin 3\theta \end{array}\right].
\end{equation}
If we decompose the deflections into the radial deflections
\begin{equation}
  \hat{e}_r \cdot \vec{\alpha}_2 = 
      -r \left[\gamma(r)+\Gamma(r)\right]\cos 2\theta
\end{equation}
and the tangential deflections
\begin{equation}
  \hat{e}_\theta \cdot \vec{\alpha}_2 = 
      r \left[\gamma(r)-\Gamma(r)\right]\sin 2\theta
\end{equation}
we can see that to (lowest order), a model must have two angular degrees of 
freedom in order to fit an Einstein ring, the internal shear $\Gamma_E$
and the external shear $\gamma_E$ at the Einstein ring.  Alternatively,
the overall ellipticity of the ring is set by $\gamma_E+\Gamma_E$ while
the detailed shape depends on $\chi_E=\Gamma_E/\gamma_E$.  Time delays
also depend on having the correct value of $\chi_E$ (see \citealt{Kochanek2006}).  
Like the monopole,
there are then higher order, sub-dominant terms (gradients of the quadrupole 
at the ring,
deviations of the octopole from the predictions of whatever model is producing
the quadrupole and so forth) even before considering the additional degrees
of freedom associated with variable axis orientations.  

An ellipsoidal model has, however, only one angular degree of freedom,
the axis ratio $q=1-\epsilon$.  Once the axis ratio is chosen, the ratio
$\chi_E = \Gamma_E/\gamma_E$ is fixed, so an ellipsoidal model will
only be able to fit Einstein rings produced by models with 
the same $\chi_E$.  At least for the low ellipticities used in the numerical
models we consider later in the paper, $\chi_E$ is independent of the 
actual value of $\epsilon$.  For higher ellipticities, there would be
non-linear corrections in $\epsilon$ to the ratio $\chi_E$.  

The right panel of Fig.~\ref{fig:models} shows contours of $\chi_E$ for
the softened power-law models.  The singular models have 
$\chi_E=(n-1)/(5-n)$, so the $n=2$ SIS has $\chi_E=1/3$.
Like $\xi_2$, the general expression for $\chi_E$ is analytic but
too long to be worth presenting.  Although the general morphology
of the $\chi_E$ and $\xi_2$ contours are similar, they do not 
track one another in detail.  Hence, the $\chi_E$ values of the 
models matched in $\xi_2$ (the open and closed point pairs) 
lie on different $\chi_E$ 
contours. A model at an open point will fail to provide a
good fit to the angular structure of an Einstein ring produced
by a model at the associated closed point.

This inability to simultaneously match $\xi_2$ and $\chi_E$ is 
the reason that \cite{Millon2019} find such large likelihood 
differences between mass models, not that Einstein rings have
any great ability to constrain radial mass distributions. In
\S\ref{sec:radial} we demonstrate that even large numbers 
of constraints in a fairly thick annulus around $R_E$ determine
$\xi_2$ and basically nothing else.  In \S\ref{sec:angular}
we reproduce the large likelihood differences found by \cite{Millon2019}
when trying to model an Einstein ring produced by one ellipsoidal
model with an ellipsoid having a different radial mass profile. 

However, no realistic lens model consists only of a single ellipsoid.  
A barely realistic model includes an external shear $\gamma_0$, in which
case $\chi_E = \Gamma_E/(\gamma_E + \gamma_0)$ for an (anti)aligned
external shear.   By appropriately adjusting the external 
shear, the $\xi_2$ and $\chi_E$ values of the input and output 
models can be matched simultaneously.  Thus, we predict, and
find in \S\ref{sec:angular},
that adding this generically required extra parameter to the angular
structure makes the huge likelihood differences between
models found by \cite{Millon2019} simply vanish.   

\section{Constraining the Radial Mass Distribution}
\label{sec:radial}

In this section we consider only circular lenses and so simply
solve the one-dimensional lens equations in a stand alone
program.  In all the models, we make the Einstein radius
$R_E \equiv 1$.  We set the ratios of the other scales 
(like $s$) to closely match the dimensionless scale ratios
of \cite{Millon2019}, although as relatively round numbers.
For each test, we generate the images for either 4 or
50 multiply imaged sources and then model them without
adding any noise.  

We fit only the image positions, scaling the goodness of fit 
statistic $\chi^2$ assuming astrometric errors of $\sigma = 0.004R_E$. 
For $R_E=1\farcs0$, this uncertainty of 0\farcs004 is 10\% of an HST WFC3/UVIS 
or ACS pixel and 3\% of a WFC3/IR pixel.  The positions of the
point-like quasar images can be measured somewhat better, although
in the CfA-Arizona Space Telescope Lens Survey (CASTLES, e.g., \citealt{Lehar2000})
we generally limited our astrometric uncertainties to about this
scale ($0\farcs003$) due to systematic differences from different 
PSF models, extended emission, pixelization and millilensing. 
The effective astrometric accuracy associated with the extended
emission of an Einstein ring, which is what we are mimicking using
large numbers of multiply imaged sources, will be lower because 
the emission is smooth.  In any case, the changes in likelihood
between models will be representative of any error model.
Because we added no noise, a fit using the 
input model yields $\chi_{in}^2=0$, so the 
likelihood ratio between the input model and a fit with
an alternative model leading to a fit statistic $\chi^2$ is simply 
$\exp(-\chi^2/2)$.  

\begin{figure*}
\centering
\includegraphics[width=0.80\textwidth]{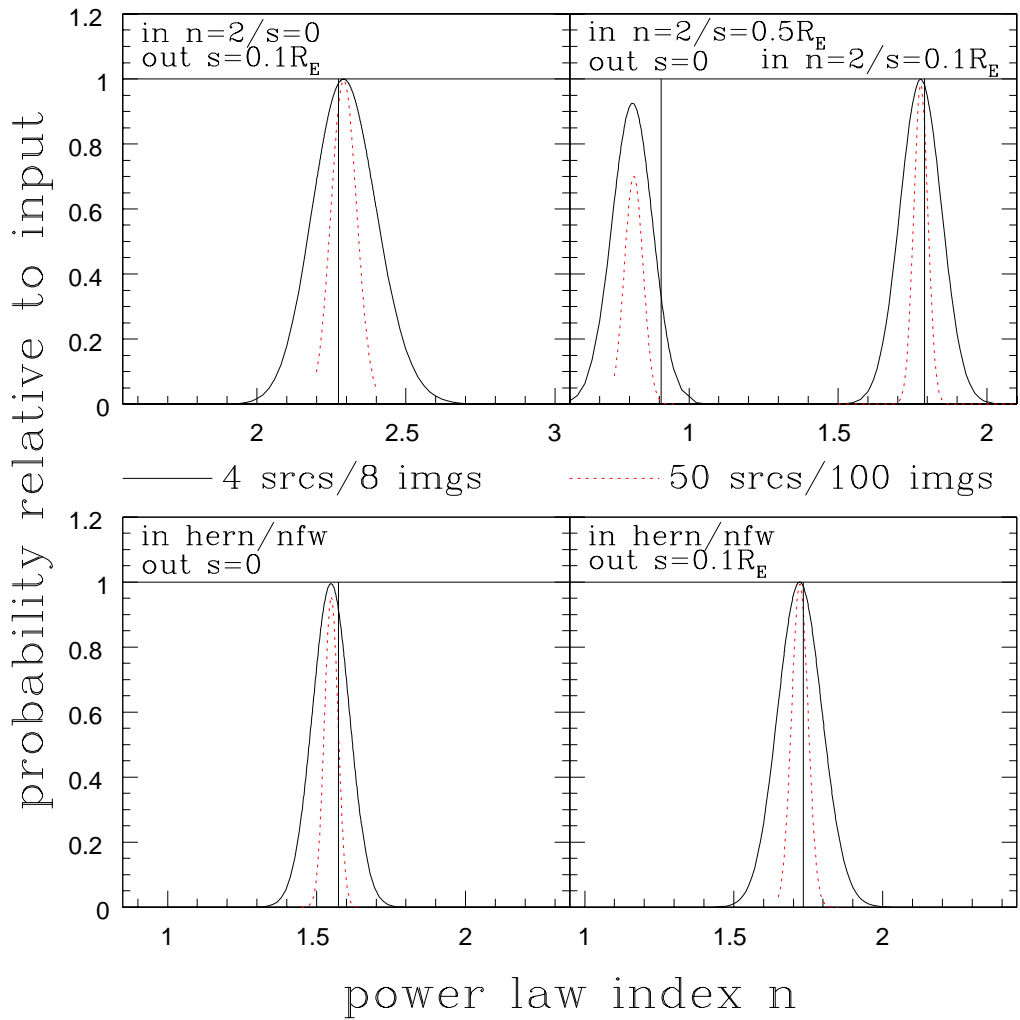}
\caption{ 
  Likelihood ratios relative to the true model as a function
  of the power law exponent $n$ for either 4 lensed sources (black
  solid) or 50 lensed sources (red dashed).  The vertical line
  indicates the value of $n$ predicted by matching the values of
  $\xi_2$.  The top left panel is for a $n=2$ singular isothermal 
  sphere modeled by a cored power law with $s=0.1R_E$.  The top
  right panel is for either a $n=2$, $s=0.5 R_E$ or a $n=2$, $s=0.1 R_E$
  cored power law modeled
  by a singular power law.  The lower panels are for a Hernquist
  plus NFW lens modeled by either a singular (left) or $s=0.1 R_E$
  cored power law (right).  
  }
\label{fig:circ}
\end{figure*}

For the four source case, we placed the outer images of each
image pair at $r_{out}=1.01$, $1.1$, $1.2$ and $1.3 R_E$.
and then solve for the position of the inner image to produce
fake image data.  For the fifty source case, we randomly selected
radii for the outer image as $r = R_E (1 + 0.3 P^{1/2})$, which roughly
corresponds to uniformly sampling a disk centered behind the lens.
In all of these models we fix
the position of the lens galaxy to its true position.  Allowing
the position of the lens galaxy to shift will lead to a further
reduction in the ability to differentiate models, but it will be
a small correction for models with a large number of sources and images.

The first two cases considered by \cite{Millon2019} model a 
lens produced by a singular isothermal sphere (SIS, $n=2$, $s=0$)
with the general power law lens.  As noted earlier, the
singular models have $\xi_2 = 2(n-2)$, so this input model has
$\xi_2 = 0$.  As a function of $n$ we can determine the core
radius $s$ needed to have $\xi_2=0$, finding that there are 
no solutions for $n<2$, and that the necessary core radius 
then increases with $n$, starting from $s=0$ at $n=2$.  So, for example,
a lens with $n=2.274$ and $s=0.1R_E$ should be virtually 
indistinguishable from the input model (see Fig.~\ref{fig:models}).  

The results for fitting a $n=2$, $s=0$ model with $s=0.1R_E$
models are shown in the upper left panel of Fig.~\ref{fig:circ}.
The cored model fits
either case almost perfectly, and almost exactly at the
power law index predicted to match the values of $\xi_2$.
The two models would, however, yield significantly 
different estimates of $H_0$ since the singular $n=2$
model has $\kappa_E=1/2$ while the cored model has
$\kappa_E=0.44$,  Note, however, that the inability
to distinguish the two models is due neither to using a
narrow annulus (the width is 60\% of $R_E$) nor due to
having few constraints (there are 50 image pairs). 

The second set of models considered by \cite{Millon2019} generate
a lens with a $n=2$ cored power law and then fit it using a 
singular model.  In round numbers, the input model has
$n=2$ and $s=0.5R_E$.  As noted earlier, this is
a somewhat pathological model due to the large core. Its
power-law match at $\xi_2=-2.189$ has $n=0.906$, which is also
somewhat pathological because the model has a radially 
increasing surface density. Because of the large radial
critical curve of the input model, we had to move the 
outermost image radius from $1.3R_E$ to $1.25 R_E$ to keep
all the sources multiply imaged.  Nonetheless, we can still
check the mathematical statement that the models should be
virtually indistinguishable.  As we see in the upper
right panel of Fig.~\ref{fig:circ}, the match is not quite
as good as in the first example.  The model is somewhat 
offset from the value of $n$ predicted by matching the values
of $\xi_2$ and the likelihood ratios are modestly different
from unity.  Still, even with 50 multiply imaged sources,
the likelihood ratio is $0.7$, which is not very significant. 

As a more realistic version of this test, we used $n=2$ and
$s=0.1R_E$ for the input model, which now has a demagnified
central image and allows us to move the outermost images
back to $1.3 R_E$.  As shown in Fig.~\ref{fig:models} the
predicted power law match in $\xi_2$ has $n=1.791$, and 
this model provides an essentially perfect fit whether 
we use four or fifty multiply imaged sources, as also 
shown in the upper left panel of Fig.~\ref{fig:circ}.

The final set of models considered by \cite{Millon2019}
combine a \cite{Hernquist1990} and a NFW (\citealt{Navarro1997})
profile to generate the lens.  To match their first such model, 
we used a Hernquist scale length of $s=1.5R_E$, an NFW 
scale length of $a=30R_E$ and normalize the models to have
$\kappa=3$ at $r=0.1R_E$ so that the density profile closely
matches that in \cite{Millon2019}.  The resulting model
has $\xi_2=-0.852$ which corresponds to $n=1.574$ for
a pure power law model and $n=1.734$ for a power law
model with a $s=0.1R_E$ core radius.  As shown in the
lower panels of Fig.~\ref{fig:circ}, the models matched 
in $\xi_2$ again provide near perfect fits for both 4 
and 50 multiply imaged sources.  The last model considered
by \cite{Millon2019} chooses parameters for the Hernquist
and NFW profiles to produce a density distribution that 
very closely mimics the singular $n=2$ power law model, 
so there is nothing new to be tested in this case.

Not surprisingly, mathematics works, and it is very difficult
to distinguish radial mass distributions matched in $\xi_2$
even with very large numbers of lensed images assumed to 
have very well-measured positions.  With even a little 
more freedom in the radial mass distribution, the small
remaining likelihood differences would be relatively easy 
to eliminate.  In short, multiple lensed sources and 
Einstein rings basically constrain nothing about the 
radial mass distributions other than $R_E$ and $\xi_2$. 

\section{How the Angular Tail Wags the Radial Dog}
\label{sec:angular}

\cite{Millon2019} argue that the reason they can distinguish radial
mass distributions is because of the large numbers of constraints 
supplied by the Einstein ring images of the hosts.  As we demonstrated
in the previous section, even large numbers of radial constraints 
spanning a fairly broad annulus around the Einstein radius cannot 
distinguish radial mass distributions with the same value of $\xi_2$.
Einstein rings do, however, provide a huge number of constraints on
the {\it angular} structure of the gravity.  This can be seen both
in the theory of Einstein ring formation (\citealt{Kochanek2001}) and in the ability
of the rings to constrain deviations in the gravity from ellipsoidal
(e.g., \citealt{Yoo2005}, \citealt{Yoo2006}).  
The \cite{Millon2019} simulations assumed ellipsoidal models with
no external shear, so they had very limited degrees of freedom in
the angular structure of the gravity.  

For each input mass distribution, we first model it as an ellipsoid
without any external shear, and then as an ellipsoid plus an
(anti)aligned external shear.  We show the results for both 4 and
50 multiply imaged sources to illustrate the consequences of adding more
and more constraints on the angular structure for the inferred
likelihood ratios of the radial structures.  
Then at the end of the section
we consider models with more complex angular structures like the
misaligned \cite{Hernquist1990} plus NFW (\citealt{Navarro1997})
models in \cite{Millon2019}.

\begin{figure}
\centering
\includegraphics[width=0.45\textwidth]{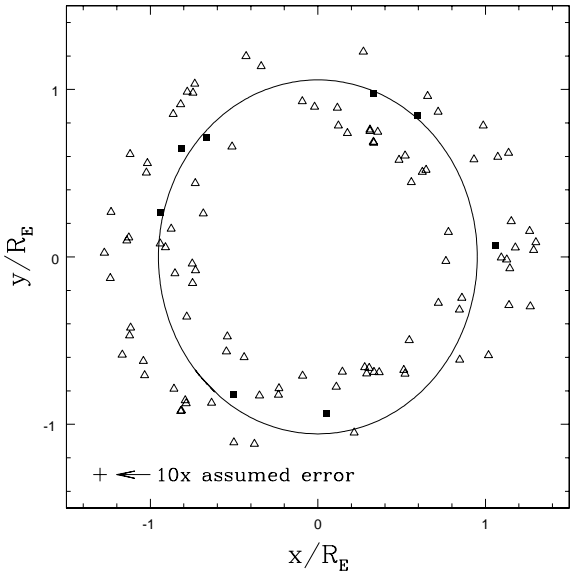}
\caption{
  The distribution of the 104 lensed images used for the $n=2$,
  $s=0$ ellipsoidal input model.  There are two sets of four
  images (filled squares), 48 sets of two images (open
  triangles) and $104$ images in total.
  The curve shows the location of the tangential
  critical line, and this singular lens has no radial critical
  line.  The sources used to generate the images were randomly
  distributed over a disk of radius $0.3 R_E$ on the source plane.  
  The cross in the lower left corner is ten times larger than the
  assumed astrometric errors of $0.004 R_E$.
  }
\label{fig:images}
\end{figure}

For the numerical models in this section we use {\tt lensmodel} (\citealt{Keeton2001},
\citealt{Keeton2011}) to generate and fit the test cases.  For the four source case
we place sources at radii of $0$, $0.1$, $0.2$ and $0.3R_E$ on the source plane and
at a random angle.  For the 50 source case we randomly
distributed the sources uniformly over a source plane region of radius $0.3R_E$.  The 
angular positions are chosen randomly.  The \cite{Millon2019} models all have
axis ratios of $q \simeq 0.9$, so we simply set $q=0.9$.  These models are 
nearly circular, so they produce very few four image systems.  The four
image cross section of an elliptical lens is of order $(\epsilon R_E/3)^2$ where
$q=1-\epsilon$ and $\epsilon/3$ is roughly the ellipticity of the potential, so
only $\sim 1\%$ of the region inside a source radius
of $0.3 R_E$ will produce four images.  Fig.~\ref{fig:images} shows the 
104 images from 50 multiply imaged sources (i.e., two sources produced
four images, the rest two images) for the first input case we consider 
with $n=2$ and $s=0$.  The symbols used in the plot are roughly ten times
larger than the assumed astrometric uncertainties of $0.004 R_E$.

For the basic models we use a single axis ratio for the input models and 
align the models with the coordinate axes.  We then model the system holding the
lens position fixed and forcing the model ellipsoid and shear to be 
(anti)aligned with the same axes. The fits would improve if these were
allowed to vary.
In their input \cite{Hernquist1990} plus NFW (\cite{Navarro1997} models,
\cite{Millon2019} allow them to have slightly different axis ratios and
to be slightly misaligned.  Obviously, a single ellipsoid fit to such a model 
has too few degrees of freedom in its angular structure, so we will return
to allowing these extra degrees of freedom in the input model after first
considering the simple case where the two profiles are aligned and have the
same ellipticity.

We do not include the
$n=2$, $s=0.5 R_E$ input model in this section, as {\tt lensmodel} has difficulty
finding solutions for the matched $n \simeq 0.9$ power-law with a radially increasing surface 
density. The difficulties probably arise because this model is so close to the degenerate $n=1$ 
constant surface density model and it is a regime where there was no physical need to
ever make {\tt lensmodel} work reliably.
We could compute a goodnesses of fit using {\tt lensmodel}'s ``source plane'' fit statistic (which is
really the position mismatch on the source plane locally corrected for image magnifications),
but not for the true ``lens plane'' fit statistic.  The qualitative results for the
``source plane'' fit statistic agree with those for the other cases but the quantitative
reliability of the results is unclear.   Since both the input and output
models are unrealistic, we study only the $n=2$, $s=0.1 R_E$ case we introduced in \S3.

\begin{figure*}
\centering
\includegraphics[width=0.80\textwidth]{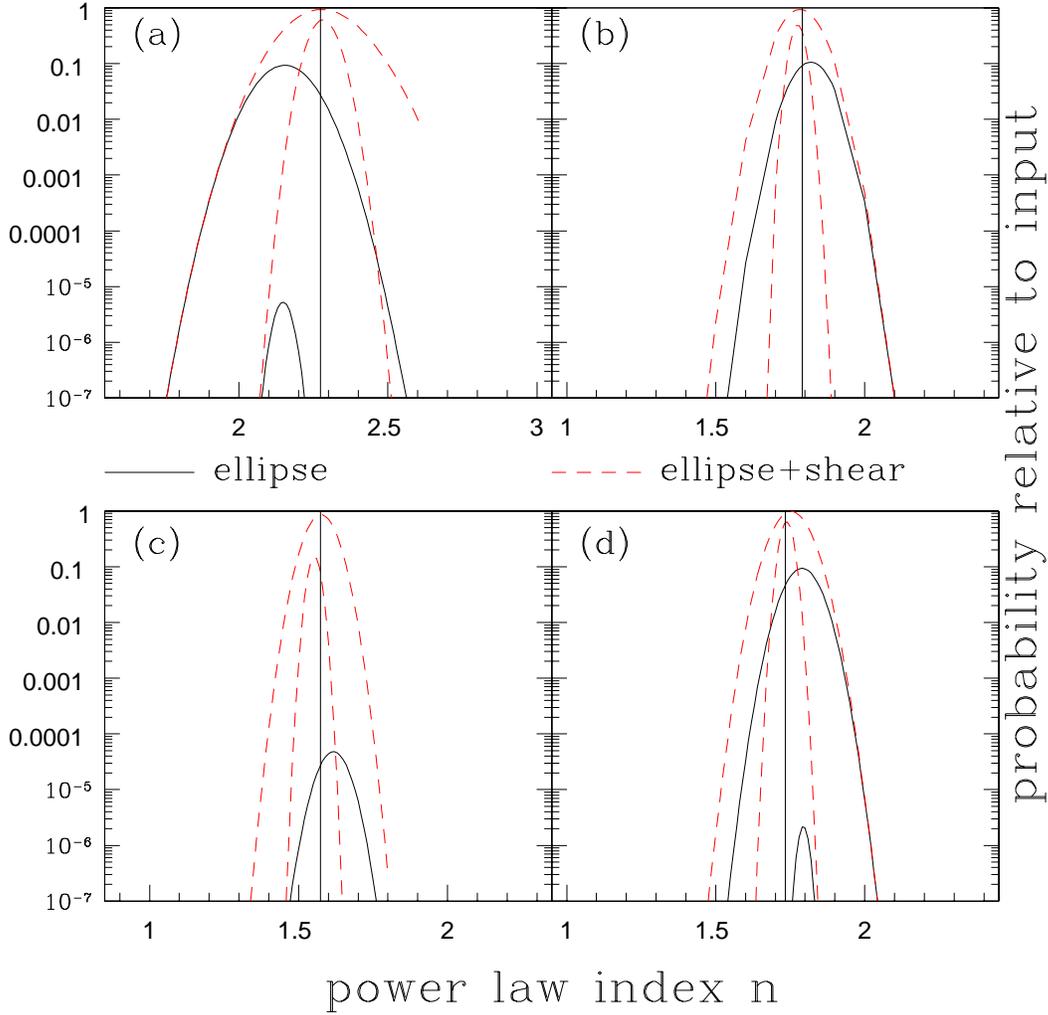}
\caption{
  Likelihood ratios relative to the true ellipsoidal model as a function
  of the power law exponent $n$ for either a purely ellipsoidal model
  (black solid) or an ellipsoid plus an (anti)aligned shear (red
  dashed).  
  The top left panel (a) is for a $n=2$ SIS sphere modeled by a cored power 
  law with $s=0.1R_E$.  The top
  right panel (b) is for a $n=2$, $s=0.1 R_E$ cored power law modeled
  by a singular power law.  The lower panels are for a Hernquist
  plus NFW lens modeled by either a singular (left, (c)) or $s=0.1 R_E$
  cored power law (right (d)).  
  In each case, the higher likelihood ratio model has 4
  multiply imaged  sources and the lower likelihood one has 50. 
  In some cases, the 50 source purely ellipsoidal model has a peak
  likelihood too low to appear.  
  }
\label{fig:ell}
\end{figure*}

We start with the input SIS ($n=2$, $s=0$) input model, where the images for
the 50 source realization are shown in Fig.~\ref{fig:images}.  As shown in
Fig.~\ref{fig:models}, the $s=0.1 R_E$ model matched in $\xi_2=0$ has $n=2.274$,
but this model differs in its angular structure $\chi_E$ from the input
model.  The top left panel of Fig.~\ref{fig:ell} shows the results. With four multiply imaged
sources, the log likelihood ratio relative to the input model for four sources
is $\sim -1.0$~dex, while for 50 sources it is $\sim -5.3$~dex, which \cite{Millon2019}
interpret as successfully distinguishing between the radial mass distributions.
The best models are also shifted away from the value of $n$ which would match the
input value of $\xi_2$ towards the $n$ of the input model.  This allows the
model to come closer to the angular structure of the input model. However,
if we now add an (anti)aligned shear, the log likelihood ratios
become $-0.03$ and $-0.22$~dex, respectively, and the models are practically 
indistinguishable ($-0.22$~dex corresponds to $\Delta\chi^2=1$).  
The best value of $n$ is also now centered on the value
predicted from matching the values of $\xi_2$.

The top right panel of Fig.~\ref{fig:ell} shows the results for modeling the
$n=2$, $s=0.1 R_E$ softened power law model with a singular power law.  For the
purely ellipsoidal model, the log likelihood ratios are $-1.0$ and 
$-7.4$~dex, respectively, where the likelihood curve for the 50 source
case does not even appear in the figure despite the dynamic range. 
However, with the addition of the anti(aligned) shear, the likelihood
ratios drop to $-0.03$ and $-0.3$~dex, again making the models practically
indistinguishable.  

Finally, Fig.~\ref{fig:ell} shows the results for the \cite{Hernquist1990}
plus NFW (\citealt{Navarro1997}) input models, where for a first test we
gave the two profiles the same $q=0.9$ axis ratio and the same major axis
position angle.  If we first consider the singular models without any
external shear, the likelihood ratios are enormous at $-4.3$ and $-20.1$~dex
for four and 50 multiply imaged sources, respectively.  When we add a
(anti-)aligned shear, the likelihood ratios drop to $-0.1$ and $-0.8$~dex,
respectively.  Similarly, the $s=0.1R_E$ models have poor fits as only
ellipsoids (likelihood ratios of $-1.0$ and $-5.7$~dex) and quite good
fits as ellipsoids plus an external shear (likelihood ratios of $-0.0$
and $-0.2$~dex).

In practice, the combined \cite{Hernquist1990} and NFW (\citealt{Navarro1997})
models used by \cite{Millon2019} used slightly 
different major axis position angles ($\Delta PA = 2.1^\circ$ for their model \#5)
for the two components.  It is unclear why adding additional angular 
structure and then fitting a single ellipsoid was viewed as a test of
recovering the radial mass distribution.  As an additional experiment
we generated a similar model, using $q=0.9$ for both components and an
axis shift of $\Delta PA = 2^\circ$ and then fit it with both the
$s=0$ and $s=0.1 R_E$ power law models allowing the orientations of 
both the ellipsoid and the shear to vary.  Considering only the 50
source models, the best fit ellipsoid with $s=0$ ($s=0.1R_E$) 
had a likelihood ratio of $-0.80$~dex at $n=1.45$ ($-0.19$~dex at $n=1.26$),
which is surprisingly good given that the model simply cannot fully
reproduce the angular structure of the input model.  Nonetheless,
the additional angular structure from having two misaligned model 
components worsens the fits compared to the models where the two
components were kept aligned.  This increases the apparent likelihood
difference between the radial mass distributions, but it is a again false
inference created by the assumed angular structures rather than an ability
to discriminate the radial mass distributions.

\section{Discussion}

It is true, as \cite{Millon2019} argue, that Einstein ring images of host galaxies
(or equivalently large numbers of multiply imaged sources as we use here) 
provide a large number of constraints on a lens model.  It is, however,
exceedingly dangerous to impose large numbers of constraints on lens models
with insufficient degrees of freedom.  This has been discussed many times
in the context of the radial mass distribution (\citealt{Gorenstein1988},
\citealt{Kochanek2002}, \citealt{Kochanek2006}, \citealt{Schneider2013},
\citealt{Wertz2018}, \citealt{Sonnenfeld2018}, \citealt{Kochanek2019}).  
As we demonstrate in \S2, Einstein rings are not very good at 
discriminating between radial mass distributions -- they will simply
identify models with the same $\xi_2$ and little else, as we argued
in \cite{Kochanek2019}.  

Einstein rings are, however, exceedingly good at determining the angular
structure of the gravitational potential (\citealt{Kochanek2001},
\citealt{Yoo2005}, \citealt{Yoo2006}).  If there are insufficient 
degrees of freedom in the allowed angular structure of the models, this
will drive the selection of the radial mass distribution and may still
lead to a poor fit.  In their models to rebut \cite{Kochanek2019},
\cite{Millon2019} find enormous likelihood ratios between the models
and interpret this as being able to distinguish the radial mass distributions.
However, as we show in \S2, the results were entirely driven by 
restricting the mass models to be ellipsoids without an external shear.
When we take the same models and include an external shear, the likelihood
differences nearly vanish, and there is essentially no ability to 
distinguish the radial mass distributions even when using 50 
multiply imaged sources with positions measured to $0\farcs004$
for an Einstein radius of $R_E=1\farcs0$.
By adding a few additional degrees of freedom to either the radial or 
angular structure of the mass model, one could reduce the rather 
statistically insignificant residual differences still further.  

The only way to be certain that the angular information is not driving
an apparent ability to discriminate between radial mass distributions
(and hence the value of $H_0$)
is to ensure that the angular structure has all the physical degrees of freedom
of real galaxies.  All models of real lens systems include external 
shears, one reason that the actual H0LiCOW (e.g., \citealt{Wong2019}) lens models do not find
likelihood ratios between monopole models nearly as large as in 
\cite{Millon2019}.  However, even an ellipsoid plus an external shear
clearly has too few degrees of freedom to have any confidence that
a statistical difference between two models for the monopole is 
being driven by an actual ability to distinguish the monopoles,
rather than it being an illusory distinction driven by assumptions in
the angular structure.  

Physically, we know galaxies are minimally comprised of both a stellar
component and a dark matter component and that these will have different
ellipticities and can be modestly misaligned.  But it is much more 
complex than that, because we also know that they can show ellipticity
gradients, axis twists, and deviations from ellipsoidal isodensity
contours (e.g., ``boxy'' or ``disky'' isophotes).  All of these
complications steadily decouple the angular structure of the gravity
from the monopole of the gravity.  Suppose, for example, that we
consider a one parameter series of monopoles, like the power law models
and generate a lens with $n=2$ but with an ellipticity that increased
with radius.  If we model this lens with a simple ellipsoid, the strong
constraints of an Einstein ring will disfavor $n=2$ because it is
producing too little exterior shear as compared to interior shear.
The models will be driven to a shallower radial mass profile (smaller
$n$) because, with more mass outside the Einstein ring, the model
can increase the exterior shear relative to the interior shear.  This
of course then produces a bias on any estimate of $H_0$.  

In short, without models that include many more angular degrees of freedom, it 
would be best to simply not include the constraints from Einstein rings.  
It is also another reason that the power law models should simply
be abandoned.  Not only do they have a one-to-one mapping between 
$\xi_2$ and $\kappa_E$ that will systematically underestimate the 
uncertainties in the convergence at the Einstein radius (and $H_0$),
but they also have a far too little freedom in their angular structure,
particularly if you are fitting Einstein rings.    
Even if you had some legitimate basis (which you do not) to ignore ellipticity gradients, 
axis twists and deviations from ellipsoidal isodensity contours,
you still have a stellar mass distribution and a dark matter distribution 
which are essentially guaranteed to have different ellipticities and even
this most basic property of real galaxies cannot be captured by the power-law
models.  The lack of an independent
parameter for the difference in ellipticity on small (stars) and large (dark matter)
scales, is essentially another ``knob'' like the shear we considered
here. But it is more general because when combined with the freedom from the 
external shear, it provides more degrees of freedom
for higher order effects like the gradients in the angular structure 
at the Einstein ring and the ability to adjust the higher order
multipoles that can modify the structure of Einstein rings while 
keeping the quadrupole structure fixed.   

The statistical approach used by \cite{Millon2019} also has a problem
in that it penalizes models for including degrees of freedom which
must be present in real galaxies.  \cite{Millon2019} use the 
Bayesian Information Criterion (BIC), 
\begin{equation}
  BIC = k \ln n - 2 \ln L
\end{equation}
where $L$ is the model likelihood, $k$ is the number of parameters
and $n$ is the number of constraints. The BIC heavily penalizes
the introduction of new parameters when there are large numbers of 
constraints $n$, as is true of Einstein ring images. 
For example, the alternative Akaike Information Criteria,
\begin{equation}
  AIC = 2 k - 2 \ln L
\end{equation}
penalizes the addition of new parameters far less than the BIC. Viewed
as a change in a $\chi^2$ statistic ($ \ln L = -\chi^2/2$), AIC views
the introduction of a new parameter as neutral if $\Delta \chi^2=1$,
while BIC views it as neutral if $\Delta \chi^2 = (1/2)\ln n$. Philosophically,
AIC should be preferred over BIC for problems like determining $H_0$ where it
is important to avoid obtaining a precise but potentially inaccurate result.  

More deeply, however, the information criteria should only be 
applied to the introduction of new parameters for which there is a  
plausible physical reason that the parameter values are known 
{\it a priori} and so holding them fixed is a reasonable prior.  
But this simply is not true for any aspect of standard
lens models -- our {\it a priori} knowledge is that the standard
models are too simple and require additional parameters if they 
are to be realistic models of the actual mass distributions of 
galaxies.  The more complex models are intrinsically more 
probable, not less probable, than the simple models, exactly
the opposite of the assumptions of the information criteria. 
The proper way to treat these complexities is to include all 
the degrees of freedom of real galaxies but with priors on their
values (e.g., ellipticity gradients are not zero, but they are 
small, etc.).  For Einstein rings, the same issues
hold for models of the source galaxy if they are parametrized analytic
models rather than pixellated source models.  Quasar host galaxies are
no more likely to be perfect ellipsoids than lens galaxies.

Finally, as noted in \cite{Kochanek2019}, the H0LiCOW (e.g., \citealt{Wong2019})
models show too little
sensitivity to the available stellar dynamical constraints compared to expectations,
and \cite{Millon2019} document this lack of sensitivity extensively. The lack
of sensitivity to the dynamical data is not a positive aspect of the existing
models -- it is a clear proof that the mass models have too few degrees of freedom.
The lens data so tightly constrain $R_E$ and $\xi_2$ that the dynamical 
information is effectively ignored because of its larger fractional uncertainties. 
This is unfortunate, because, unlike Einstein rings, dynamical
data actually does help to constrain $\kappa_E$.  At least for the
radial mass distribution, one would actually have more reliable constraints 
on $\kappa_E$ by simply ignoring the Einstein ring and relying on the 
dynamical data.  A simple test for whether mass models have sufficient
degrees of freedom is that the they should show the expected sensitivity
to the dynamical data, namely that the fractional uncertainties in $H_0$
should be comparable to the fractional uncertainties in the velocity 
dispersion (see \citealt{Kochanek2019}).    

\section*{Acknowledgments}

The author thanks M. Millon for answering many questions.
CSK is supported by NSF grants AST-1908952 and AST-1814440.

\end{document}